\documentclass[aps,prb,preprint,showpacs,amsmath,amssymb]{revtex4}

\usepackage[dvips]{graphicx}

\begin{document}

\title{Phase lapses in scattering through multi-electron quantum dots:
Mean-field and few-particle regimes}

\author{Andrea Bertoni}
\email[e-mail: ]{bertoni.andrea@unimore.it}
\homepage[group web-page: ]{www.nanoscience.unimore.it}
\affiliation{CNR-INFM National Research Center on nanoStructures
and bioSystems at Surfaces ($S3$), Via Campi 213/A, 41100
Modena, Italy}

\author{Guido Goldoni}
\affiliation{CNR-INFM National Research Center on nanoStructures
and bioSystems at Surfaces ($S3$), Via Campi 213/A, 41100
Modena, Italy}
\affiliation{Dipartimento di
Fisica, Universit\`a di Modena e Reggio Emilia, 41100 Modena, Italy}


\begin{abstract}
We show that the observed evolution of the transmission phase through
multi-electron quantum dots with more than $\sim 10$ electrons, which
shows a \emph{universal} (i.e., independent of $N$) as yet unexplained
behavior, is consistent with an electrostatic model, where
electron-electron interaction is described by a mean-field
approach. Moreover, we perform exact calculations for an open 1D
quantum dot and show that carrier correlations may give rise to a
\emph{non-universal} (i.e., $N$-dependent) behavior of the
transmission phase, ensuing from Fano resonances, which is
consistent with experiments with a few ($N < 10$) carriers. 
Our results suggest that in the universal regime the
coherent transmission takes place through a single level while
in the few-particle regime the correlated scattering state is
determined by the number of bound particles.
\end{abstract}

\pacs{73.63.Kv, 03.65.Nk, 72.10.-d  73.23.Hk}


\maketitle

\section{INTRODUCTION}
Among the experiments that exploit the coherent dynamics of carriers,
the one performed in 1997 by Schuster et al.~\cite{shuster97}, in
which the transmission phase of an electron scattered through a
quantum dot (QD) was measured, constitute an ideal test on the
validity of different theoretical models for the inclusion of
electron-electron interaction. In fact, the ability to model coherent
carrier transport experiments in low-dimensional semiconductor systems
is essential for designing possible future devices for coherent
electronics or quantum computing.  
In the experiments of Refs.~\onlinecite{shuster97} and
\onlinecite{kalish05}, two paths
are electrostatically defined in a high-mobility AlGaAs 2DEG, within a
multi-terminal setup that allows to overcome the phase-rigidity
constraint~\cite{levyyeyati95} of a two-terminal one.
Two narrowings along one of the paths
define a QD which is operated in the Coulomb blockade regime (a
different set of experiments, performed in the Kondo regime, presents
another peculiar phase behavior~\cite{ji00,silvestrov03}).  The
transmission phase across the QD is measured by an electron
interferometry technique in which electrons are emitted at a given
energy from a quantum point contact at one end of the two-path
system: When the energy corresponds to a quasi-bound level (QBL) of
the QD, a transmission resonance occurs. The depth of the QD confining
potential $V_d$ is tuned by charging a nearby ``plunger'' gate and the
transmittance, together with the corresponding phase, is obtained as a
function of $V_d$.

The process of electron scattering through the QD has been modeled by
means of a number of different approaches, ranging from multi-particle
few-sites~\cite{oreg97} to lattice~\cite{levyyeyati00} and
Hubbard~\cite{xu01} model
Hamiltonians~\cite{silvestrov00,hackenbroich01}.  Still, none of the
proposed approaches has been able to fully reproduce the main feature
of the measured transmission phase $\theta$, namely, the recurring
behavior found in the many-particle regime of the QD, where $\theta$
smoothly changes by $\pi$ on each transmission peak of the
$N$-electron system, and then abruptly drops to the initial value in
each valley between the $N$ and $N+1$ resonances, this leading to
in-phase transmission resonances. This is called the \emph{universal
behavior} since it does not depend on the charge status of the
QD. While the change of the phase at each resonance is well described
by the Breit-Wigner model, the nature of the phase drops remains
substantially unexplained.

Recently, an enhanced version of the electron interferometer
system~\cite{kalish05}, allowing for the precise control of the
number of electrons inside the QD down to zero, has been used to
measure the coherent transmission amplitude for small $N$. The
results show that when only a few electrons ($N<6$) are bound
into the QD, the universal behavior of the phase is lost, and the
phase drop occurs only for certain values of $N$.
Furthermore, it was confirmed that the measured phase evolution is
indeed related to the $N$-electron dot and not to the larger
two-path device.

The aim of the present paper is to show that the universal behavior of
the phase (large $N$) is consistent with an electrostatic
approximation, where the electron-electron interaction between the
scattered carrier and the bound ones is included as a mean Coulomb
field. This is done in Sec.~\ref{meanfield}, where the transmission
probability and phase are computed for a 2D potential representing the
QD (attached to source and drain leads) plus a ``large'' number of
bound electrons.  Furthermore, in Sec.~\ref{fewpart} we show that an
exact few-particle calculation performed on an effective 1D model of
the system leads to the appearance of both Breit-Wigner and Fano
resonances \cite{nockel94}, with continuous and discontinuous phase
evolution, respectively, consistent with the experimental findings in
the small $N$ regime. Finally, in Sec.~\ref{conclusions}, we draw our
conclusions.

\section{MEAN-FIELD APPROACH: RECURRING PHASE DROPS}
\label{meanfield}
\begin{figure}
\includegraphics[width=.8\linewidth]{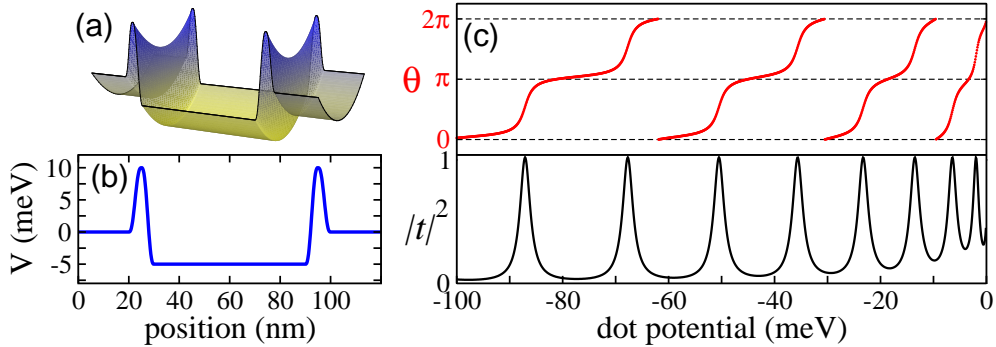}
\caption{\label{fig1}
(color online).
The adopted 2D potential profile (a) consisting of a harmonic
potential with level spacing of 30~meV in the transverse direction
and a double barrier along the propagation direction (b).
Since no bias is applied the Fermi levels in the source and the drain
coincide, and are taken to be zero.
Single-particle transmission probability and phase (c) are
shown, as a function of the QD potential. The energy
of the incoming carriers is 1~meV.
}
\end{figure}

Let us resume the expected phase evolution for a single electron
crossing an empty QD. We do so for a specific 2D potential $V_s$
[Figs.~\ref{fig1}(a) and \ref{fig1}(b)] which mimics the one generated
by the surface metallic gates in the 2DEG of the devices of
Refs.~\onlinecite{shuster97} and \onlinecite{kalish05}.
Along the propagation direction [Fig.~\ref{fig1}(b)], we take two
smoothed barriers (with a maximum height of 10~meV and a maximum width
of 10~nm) that connect a 60~nm flat negative region which mimics the
QD potential which is tuned in the simulations; along the transverse
direction, we consider a harmonic confinement with
$\hbar\omega=$30~meV. In the setup used in
Refs.~\onlinecite{shuster97} and \onlinecite{kalish05}, no bias is
applied between the QD source and drain leads since the coherent
electron traversing the dot is emitted by a quantum point contact at
one end of the two-path system (not included in our
simulations). Accordingly, we keep the Fermi energy of the two leads at
zero potential and fix the energy of the incoming electron. Material
parameters for GaAs have been used. The open-boundary single-particle
2D Schr\"odinger equation has been solved by using the \emph{quantum
transmitting boundary method}~\cite{kirknerqtbm} in a
finite-difference scheme.  Figure~\ref{fig1}(c) shows the transmission
probability and phase as a function of the QD potential. As the QD
potential is varied, the incoming electron comes into resonance with
higher single-particle QBLs. At each resonance peak, the phase
increases by $\pi$ in agreement with the Breit-Wigner model, while it
is substantially constant in the low-transmission valleys.

We show next that when the \emph{mean} Coulomb field of electrons
that populate the QD is taken into account, the behavior of the
transmission phase shows the observed drops. In our model, the
maximum of a transmission peak corresponds to the alignment of the
energy of the scattered electron with the energy of a QBL of the
mean-field potential. When the energy bottom of the QD is further
lowered and the alignment is lost, the transmission probability
decreases until the QBL becomes a genuine bound state, i.e., its
energy falls below the Fermi energy, and it is occupied by an
additional electron. The new mean-field potential has the QBL of the
previous resonance shifted by the addition energy and, after a
further lowering of the QD potential, it produces another resonance.
This phenomenon, that is essentially a Coulomb blockade effect, is
repeated each time a carrier is added to the QD. As the mean fields
produced by $N$ or $N+1$ electrons are very similar in the large $N$
regime, the QBL that generates the resonances and the corresponding
transmission phase is always the same at each peak, with an abrupt
drop each time a new electron occupies a bound state of the QD.

We now apply our model to a QD with the structure potential $V_s$ of
Figs.~\ref{fig1}(a) and \ref{fig1}(b).  In order to estimate the QD
electrostatic potential we first solve the closed-boundary
Schr\"odinger equation then add the field generated by an electron in
the ground state $\psi_1$, namely
\begin{equation} \label{eqpot2d}
V_1(x,y)= \frac{e^2}{4\pi\epsilon} \int dx' dy'
\frac{|\psi_1(x',y')|^2 e^{-r/\lambda_D} } {r}
\end{equation}
with $r=\sqrt{(x-x')^2 + (y-y')^2 + (d/2)^2}$ and where $d=1$~nm
represents the thickness of the 2DEG and $\lambda_D=30$~nm is the
Debye length~\cite{nota1}. The Fermi levels of the source and drain
leads are fixed, i.e., we neglect the effect of the charge inside the
QD on the leads. We compute the ground state of the new potential
$V_s+V_1$ and we repeat the whole procedure until we reach a number
$N$ of bound particles for which the potential $V_s+V_1+\dots+V_N$ has
an unbound (positive energy) ground state.  Then we compute the 2D
scattering state for an incoming electron with the boundary conditions
already described for the single-particle calculation.  For simplicity
the bound states are calculated in a finite domain by solving the
closed-boundary Schr{\"o}dinger equation. This leads to a shift in the
energy of the bound states that has no effect on the qualitative
results of the present work, i.e., the phase drops between the
transmission resonances.

\begin{figure}
\includegraphics[width=.8\linewidth]{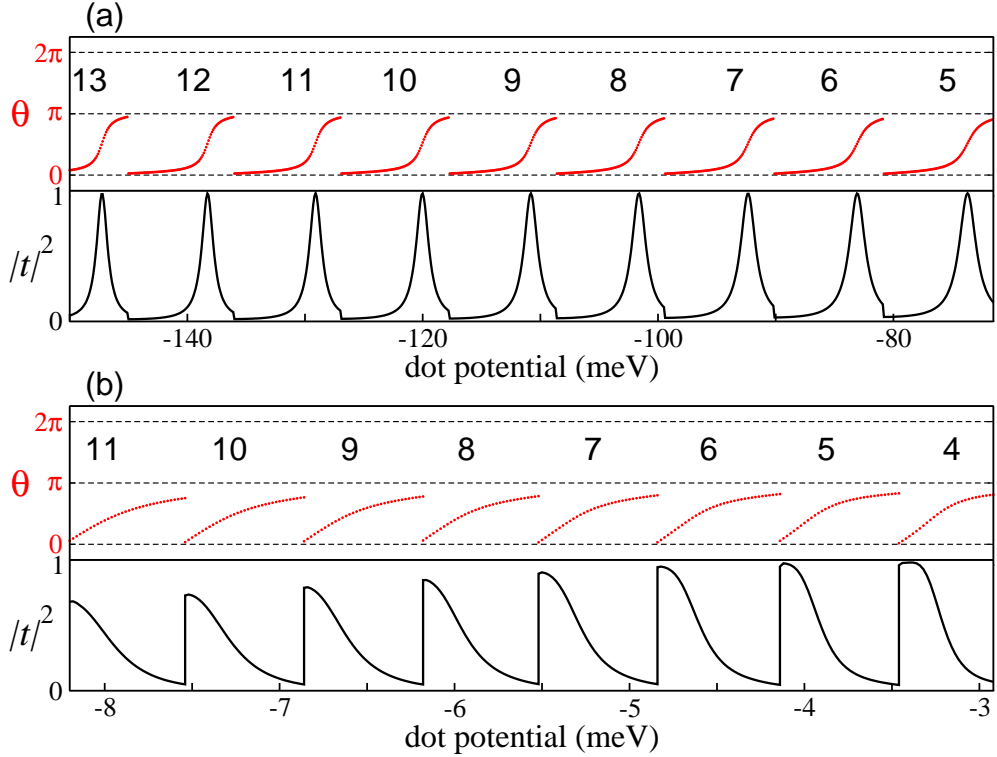}
\caption{\label{fig2}
(color online).
(a) Transmission probability and phase for an electron scattered by
the potential corresponding to the sum of the structure potential
described in Fig.~\ref{fig1} and the mean field generated by $N$ bound
electrons. The numbers indicate the value of $N$ at each transmission
peak: as the QD potential $V_d$ decreases $N$ increases. (b)
Transmission probability and phase are shown for a system similar to
the one of Fig.~\ref{fig1} whose parameters are tuned in order to
match the energy levels of Ref.~\onlinecite{kalish05}, namely: kinetic
energy of the scattered electron $\approx 20$~$\mu$eV, charging energy
$\approx 1$~meV, coupling of the QD with the leads $\approx
200$~$\mu$eV, and difference between the first two single-particle
QBLs $\approx 500$~$\mu$eV. The above values are obtained with a
$100$~nm well and two $50$~$\mu$eV barriers $4$~nm wide in the
longitudinal direction and a harmonic confinement with
$\hbar\omega=1$~meV in the transverse direction.
}
\end{figure}

We show two sets of calculations in Fig.~\ref{fig2}. 
In the top panel (a), the system parameters are chosen as in
Fig.~\ref{fig1} in order to obtain a clear resolution of the
resonances, although they do not correspond to the experiments in
Refs.~\onlinecite{shuster97} and \onlinecite{kalish05}.  For the chosen
parameters, the transmission occurs through the fourth excited QBL.
All resonances, corresponding to different $N$, are in phase and this
trend continues as the potential of the QD deepens, i.e., for larger
numbers of bound electrons. Note that, although the effect of the
charging of the QD is essentially classical, the transmitted electron
must be obviously modeled in a quantum approach in order to obtain the
transmission phase.

In Fig.~\ref{fig2}(b), we consider a structure with parameters closer
to those of the experimental conditions in Ref.~\onlinecite{kalish05};
in particular, the confinement potential is much weaker and the energy
of the incoming electron smaller than in Fig.~\ref{fig2}(a) (see
caption), leading to less defined resonances. The transmission phase
evolution is similar to the previous case, in spite of the fact that
differences between the two calculations are not only quantitative,
showing that the results obtained are robust against the details of
the calculation and of the system.  In particular, (1) due to the low
energy of the incoming carrier, the transmission takes place through
the ground QBL rather than an excited state; (2) the two lowest QBLs
are, for $N>4$, quasi-degenerate.\cite{nota4} The latter effect is due
to charge accumulation in the center of the QD, away from the
barriers, inducing a double-well-like profile along the propagation
direction.  In this regime, the transmission peaks corresponding to the
two lowest QBLs merge and, for each $N$, a single transmission
resonance is found that, being originated by two quasi-degenerate
states, is characterized by a phase change of $2\pi$. However, since
the trapping of an additional electron in a localized state takes
place just after the transmission maximum, the resulting phase
evolution spans only a range of $\pi$. A further effect of the charge
accumulation in the center of the QD is the decrease of the maximum
value of the transmission probability on the resonances. The above
trend is clear in the left part of Fig.~\ref{fig2}(b).  We note that
our simulations are performed at zero temperature and with an exact
energy of the incoming carriers, this leading to the steep transitions
in the transmission probability of Fig.~\ref{fig2}(b).  Such steepness
is not expected in experiments due to the uncertainty of the incoming
carriers' kinetic energy and the temperature dependence of bound levels'
occupancy.  In the simulations based on the mean-field approach, the
universal behavior, i.e. the phase drops occurring between successive
resonances, persists down to $N=0$, in contrast with experiments of
Ref.~\onlinecite{kalish05} where phase drops may or may not occur for
$N<6$. It should be noted, however, that the phase drops are a
necessary consequence of the electrostatic model employed for the
coupling between the bound and incoming electrons: Such a mean-field
picture is expected to break down at small $N$. Indeed, we show in the
following section that the inclusion of carrier correlation may give
rise to an $N$-dependent phase evolution.

To conclude our mean-field analysis we discuss the similarities
between the results hitherto presented and the ones obtained in
Ref.~\onlinecite{xu01}, also including electron-electron interaction
in a mean-field approximation.  In the above work, the lead-dot-lead
system is modeled with a cross-bar geometry and the transmission
amplitude is obtained by using the non-equilibrium Green function
approach and a Hubbard Hamiltonian.  The recurring phase drops are
found at zeros of the transmission~\cite{notaxu} and persist when the
electron-electron interaction is turned off.  While the first effect
agrees with our simulations, we find no drops in the non-interacting
case.  The difference can be explained by the different models adopted
for the dot: a 2D double-barrier structure in our case and a 1D bar
orthogonal to the lead-to-lead direction in Ref.~\onlinecite{xu01}.
This is confirmed by the further agreement between our
Fig.~\ref{fig1}(c) and a non-interacting simulation for a
double-barrier 1D structure reported in the above work.  There, the
universal behavior of the transmission phase seems induced by the
cross-bar configuration, with a single site between the two leads,
regardless of the Coulomb interaction.

\section{FEW-PARTICLE APPROACH: BREIT-WIGNER AND FANO RESONANCES}
\label{fewpart}
In order to obtain numerically the transmission coefficient of a fully
correlated system the calculation must be able to solve the
few-particle problem exactly in an open domain, a difficult task for a
general 2D potential. We therefore chose to simulate the dynamics for
two electrons in a strictly 1D quantum wire with the same profile of
Fig.~\ref{fig2}(b); the lateral extension of the wire is, however, taken
into account by an effective Coulomb potential $V_C(x) = {e^2} /
[4\pi\epsilon \left(x+d\right)] $, where the Coulomb singularity is
smoothed by a cutoff $d = 1$~nm~\cite{fogler05}. Then, we solve exactly
the few-particle open-boundary Schr\"odinger equation in the real
space, using a generalization of the quantum transmitting boundary
method mentioned above, whose general derivation is detailed elsewhere
\cite{bertoni_jce,bertoni_unpub}.
In the following, we describe it for a 1D spinless system.

Let us consider a region of length $L$, with a single-particle
potential $V(x)$, constant outside that region (leads): $V(x)=V(0)$ if
$x<0$ and $V(x)=V(L)$ if $x>L$.  Although the method is valid for the
general case we consider here $V(0)=V(L)=0$ for simplicity.  Let us
take $(N-1)$ interacting identical particles bound by $V(x)$ in its
$(N-1)$-particle ground state $\chi_0(x_1,...,x_{N-1})$. The $m$-th excited
eigenstates of $(N-1)$ interacting particles will be denoted by $\chi_m$.

Our aim is to find the correlated scattering state of
$N$-particles $\psi(x_1,\dots,x_N)$
that has the following form when the $n$-th 
(with $n \le N$) particle
is localized in the left lead, i.e., when $x_n<0$:

\begin{widetext}
\begin{eqnarray}  \label{sym0boundary}  \nonumber
\psi(x_1,...,x_n,...,x_N) |_{x_n<0}&=&
(-1)^n \bigg[ 
\chi_0(x_1,...,x_{n-1},x_{n+1},...,x_N) e^{ik^l_0x_n} +
\\ 
&& 
+\sum_{m=0}^{M_l} b^<_m \chi_m(x_1,...,x_{n-1},x_{n+1},...,x_N)
e^{-ik^l_mx_n} +
\\  \nonumber
&& 
+ \sum_{m={M_l}+1}^M b^<_m \chi_m(x_1,...,x_{n-1},x_{n+1},...,x_N)
e^{k^l_mx_n}  \bigg] ,
\end{eqnarray}
where $(-1)^n$ accounts for the wave function antisymmetry and
$k^l_m=\sqrt{|2m_eT_m|}$ represents the wave vector of the traveling
particle, with mass $m_e$, whose kinetic energy $T_m=E-E_m$ is
obtained from the total energy $E$ and the energies $E_m$ of the
states $\chi_m$; in turn $E$ can be obtained from $E=T_0+E_0$ since
the incoming-particle energy $T_0$ is the Fermi energy in the left
lead.

The first term inside the square brackets on the r.h.s. of
Eq.~(\ref{sym0boundary}) accounts for the $n$-th particle incoming as
a plane wave with $T_0$ energy from the left lead, while the other
$(N-1)$ particles are in the ground state of $V$.
The second term represents the linear combination of all the
energy-allowed possibilities with the $n$-th particle reflected back
as a plane wave in the left lead with energy $T_m$ and the dot in the
$\chi_m$ state.
The third term is analogous to the latter but accounts for the case
$E_m>E$, representing the $n$-th particle as an evanescent wave in the
left lead.  The number of bound states whose energy is lower than the
total energy, $E_m<E$, is $M_l+1$.
When particle $n$ is in the right lead, the wave function has a form
similar to Eq.~(\ref{sym0boundary}), without the incoming term since we
are considering only electrons traversing the dot from the left lead:
\begin{eqnarray}  \label{symLboundary}  \nonumber
\psi(x_1,...,x_n,...,x_N) |_{x_n>L} &=&
(-1)^n \bigg[ 
\sum_{m=0}^{M_r} b^>_m \chi_m(x_1,...,x_{n-1},x_{n+1},...,x_N)
e^{ik^r_mx_n}
\\
&& +\sum_{m={M_r}+1}^M b^>_m \chi_m(x_1,...,x_{n-1},x_{n+1},...,x_N)
e^{-k^r_mx_n} \bigg].
\end{eqnarray}
\end{widetext}

Since the number of interacting particles is $N$ and the problem is
1D, our computational domain consists of an $N$-dimensional hypercube
that we discretize with a real-space square mesh.  On the internal
points, the wave function $\psi(x_1,\dots,x_N)$ satisfies the usual
$N$-body Schr\"odinger equation
\begin{eqnarray}  \label{schreq}
\left[ -\frac{\hbar^2}{2m_e} \sum_{n=1}^{N} \frac{d^2}{dx_n^2}
+ \sum_{n=1}^{N} V_s(x_n) 
+ \sum_{n=1}^{N}\sum_{n^\prime=1}^{n} V_C(|x_n-x_{n^\prime}|)
\right] \psi(x_1,\dots,x_N)
=
E \, \psi(x_1,\dots,x_N) \, ,
\end{eqnarray}
where $V_s(x)$ is the 1D single-particle potential energy of the structure
at position $x$ and $V_C(d)$ is the effective 1D Coulomb energy of two
electrons at a distance $d$. $E$ is the total energy, defined previously.

The wave function $\psi$ has to match Eqs.~(\ref{sym0boundary}) and
(\ref{symLboundary}) on the $N$ ``left'' boundaries and the $N$
``right'' boundaries of the domain,
respectively. Equations~(\ref{sym0boundary}), (\ref{symLboundary}) and
(\ref{schreq}) are then solved together as a coupled system of $2N+1$
equations, using a finite-difference discretization for the
derivatives.  In this way, the reflection amplitudes $b^<$ and the
transmission amplitudes $b^>$ (i.e. the unknowns, together with
$\psi$) are obtained numerically.\cite{nota5} Note that the resulting
wave function is antisymmetric for any two-particle exchange since we
have imposed antisymmetric boundary conditions.

\begin{figure}
\includegraphics[width=.8\linewidth]{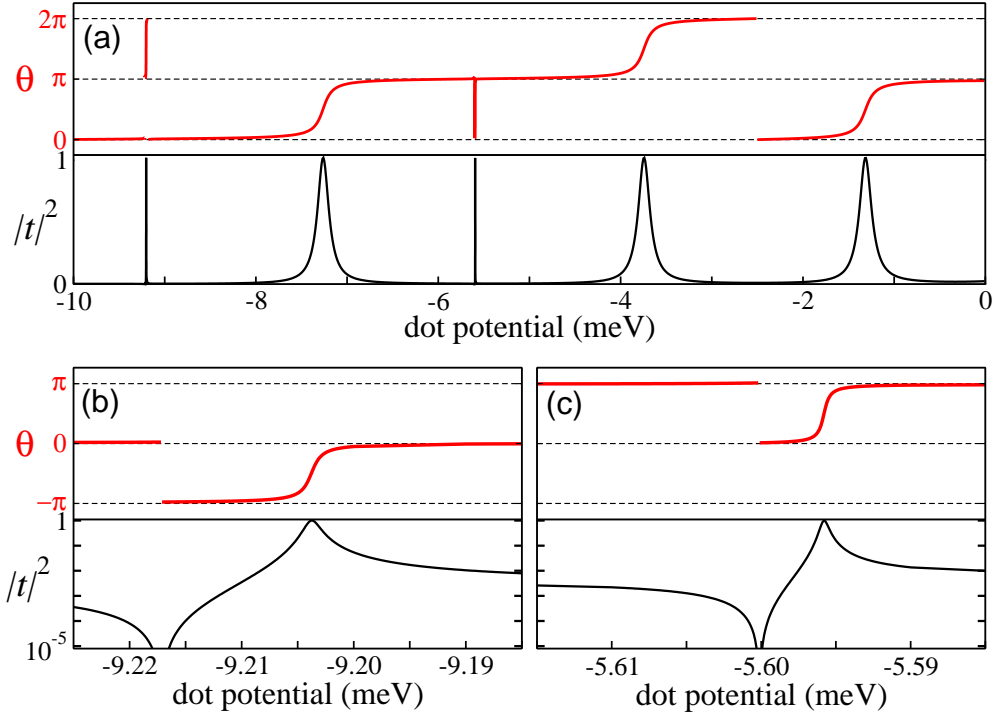}
\caption{\label{fig3}
(color online).
(a) Correlated transmission probability (bottom) and phase (top) for
an electron scattered by the potential described in Fig.~\ref{fig2}
caption when a second electron is bound in the QD. Three Breit-Wigner
and two narrow Fano resonances are present. (b) and(c) Details of the
transmission spectrum, with transmission probability in logarithmic
scale, showing the asymmetric Fano resonances and the corresponding
phase jump of $\pi$.  }
\end{figure}

Figure~\ref{fig3}(a) shows the transmission amplitude for a
two-particle correlated triplet state (bound and traveling electrons
with the same spin) as a function of the QD potential.
The energy of
the incoming electron is $T_0= 20~\mu$eV, as in the mean-field
simulations of Sec.~\ref{meanfield}.  The first five transmission
resonances are clearly visible, three of which have a Lorentzian shape
with a width of about $200$~$\mu$eV and a Breit-Wigner-type phase
evolution, similar to the ones already seen in Fig.~\ref{fig1}. The
two remaining resonances, shown in detail in Figs.~\ref{fig3}(b) and
\ref{fig3}(c) (note the logarithmic scale for the transmission
amplitude), are very narrow (few $\mu$eV) and present a typical
asymmetric Fano line shape~\cite{fano61}.  They are a signature of
electron-electron correlation and, from their small width, we deduce
that the effect of the Coulomb potential is very limited in our
model. Nevertheless, the distinctive behavior of the transmission phase
is clearly visible in the upper plots of Figs.~\ref{fig3}(b) and
\ref{fig3}(c). An abrupt phase jump of $\pi$ takes place near the
resonance, where the transmission probability vanishes. This shows
that the origin of the phase jumps detected in the few-particle
experiments may reside in correlation-induced Fano resonances. A
similar behavior, with the presence of both Breit-Wigner and Fano
resonances, showing continuous and discontinuous phase evolutions
respectively, is found in the simulation of the correlated
three-electron scattering state (not shown here). In the latter case,
the ratio of Fano resonances is larger and keeps increasing with the
number of particles.

\section{CONCLUSIONS}
\label{conclusions}
In summary, we showed that in the few-particle correlated regime, both
Breit-Wigner and Fano resonances are found, while in the mean-field
regime, a single type of resonance is present, which is repeated for
each number of bound electrons, this leading to a phase jump of $\pi$
each time a new electron enters the QD. While these results are
consistent with experiments, the microscopic nature of the recurring
phase drops in the latter regime remains unclear.  In fact, they are
well reproduced by the first of our approaches in which no quantum
correlation is present between the bound electrons and the scattered
one. On the other hand different models~\cite{hackenbroich01} that
take into account the fully-correlated dynamics of the carrier such as
our few-particle calculations cannot be applied to the many-particle
regime.  The above considerations suggest that in the latter regime,
the coherent component of the transmitted electron wave function
(i.e., the component that does not get entangled with the QD and whose
phase is detected by the interferometer) behaves as if the QD,
together with the bound electrons, was a static electric field, being
unable to discriminate between two different values of $N$. On the
other hand, when $N$ is small, the transmission phase is able to
provide a partial information on the number of electrons confined in
the QD through the character of the transmission resonance and the
possible phase lapse.  The transition regime between the many- and
few-particle conditions~\cite{taniguchi99,rontani06} needs further
analysis since it can clarify the connections between the two opposite
approaches used in the present work.

We finally note that the Fano resonances found in the 1D two-particle
scattering states are a genuine effect of carrier-carrier correlation,
in accordance to the original concept developed in
Ref.~\onlinecite{fano61}.  A more general definition is often adopted,
ascribing the Fano line shape of the transmittance to the interference
of two alternative real-space pathways. In fact, Fano resonances have
been obtained previously by means of 2D~\cite{tekman93},
multi-channel~\cite{nockel94,racec01}, and
two-path~\cite{wohlman02,fuhrer06} single-particle calculations. In
those cases, however, the ratio between the number of Fano and
Breit-Wigner resonances is not expected to vary when varying the
confinement energy of the QD, while the number of correlation-induced
Fano resonances becomes shortly dominant when moving from a low-$N$ to
a high-$N$ condition in the framework of a full few-particle
modeling.

Upon completion of this work we learned about a recent work by
Karrasch et al.~\cite{karrasch07} where the $\pi$ lapses are
also ascribed to (anti)resonances of Fano type.

\begin{acknowledgments}
We are pleased to thank M.~Heiblum, E.~Molinari and M.~Rontani for
fruitful discussions.  We acknowledge financial support by MIUR-FIRB
no. RBAU01ZEML, EC Marie Curie IEF NANO-CORR, and INFM-Cineca
Iniziativa Calcolo Parallelo 2006.
\end{acknowledgments}

\newpage

REFERENCES


\begin{thebibliography}{27}
\expandafter\ifx\csname natexlab\endcsname\relax\def\natexlab#1{#1}\fi
\expandafter\ifx\csname bibnamefont\endcsname\relax
  \def\bibnamefont#1{#1}\fi
\expandafter\ifx\csname bibfnamefont\endcsname\relax
  \def\bibfnamefont#1{#1}\fi
\expandafter\ifx\csname citenamefont\endcsname\relax
  \def\citenamefont#1{#1}\fi
\expandafter\ifx\csname url\endcsname\relax
  \def\url#1{\texttt{#1}}\fi
\expandafter\ifx\csname urlprefix\endcsname\relax\def\urlprefix{URL }\fi
\providecommand{\bibinfo}[2]{#2}
\providecommand{\eprint}[2][]{\url{#2}}

\bibitem[{\citenamefont{Schuster et~al.}(1997)\citenamefont{Schuster, Buks,
  Heiblum, Mahalu, Umansky, and Shtrikman}}]{shuster97}
\bibinfo{author}{\bibfnamefont{R.}~\bibnamefont{Schuster}},
  \bibinfo{author}{\bibfnamefont{E.}~\bibnamefont{Buks}},
  \bibinfo{author}{\bibfnamefont{M.}~\bibnamefont{Heiblum}},
  \bibinfo{author}{\bibfnamefont{D.}~\bibnamefont{Mahalu}},
  \bibinfo{author}{\bibfnamefont{V.}~\bibnamefont{Umansky}}, \bibnamefont{and}
  \bibinfo{author}{\bibfnamefont{H.}~\bibnamefont{Shtrikman}},
  \bibinfo{journal}{Nature} \textbf{\bibinfo{volume}{385}},
  \bibinfo{pages}{417} (\bibinfo{year}{1997}).

\bibitem[{\citenamefont{Avinun-Kalish et~al.}(2005)\citenamefont{Avinun-Kalish,
  Heiblum, Zarchin, Mahalu, and Umansky}}]{kalish05}
\bibinfo{author}{\bibfnamefont{M.}~\bibnamefont{Avinun-Kalish}},
  \bibinfo{author}{\bibfnamefont{M.}~\bibnamefont{Heiblum}},
  \bibinfo{author}{\bibfnamefont{O.}~\bibnamefont{Zarchin}},
  \bibinfo{author}{\bibfnamefont{D.}~\bibnamefont{Mahalu}}, \bibnamefont{and}
  \bibinfo{author}{\bibfnamefont{V.}~\bibnamefont{Umansky}},
  \bibinfo{journal}{Nature} \textbf{\bibinfo{volume}{436}},
  \bibinfo{pages}{529} (\bibinfo{year}{2005}).

\bibitem[{\citenamefont{Yeyati and Buttiker}(1995)}]{levyyeyati95}
\bibinfo{author}{\bibfnamefont{A.~L.} \bibnamefont{Yeyati}} \bibnamefont{and}
  \bibinfo{author}{\bibfnamefont{M.}~\bibnamefont{Buttiker}},
  \bibinfo{journal}{Phys.\ Rev.\ B} \textbf{\bibinfo{volume}{52}},
  \bibinfo{pages}{R14360} (\bibinfo{year}{1995}).

\bibitem[{\citenamefont{Ji et~al.}(2000)\citenamefont{Ji, Heiblum, Sprinzak,
  Mahalu, and Shtrikman}}]{ji00}
\bibinfo{author}{\bibfnamefont{Y.}~\bibnamefont{Ji}},
  \bibinfo{author}{\bibfnamefont{M.}~\bibnamefont{Heiblum}},
  \bibinfo{author}{\bibfnamefont{D.}~\bibnamefont{Sprinzak}},
  \bibinfo{author}{\bibfnamefont{D.}~\bibnamefont{Mahalu}}, \bibnamefont{and}
  \bibinfo{author}{\bibfnamefont{H.}~\bibnamefont{Shtrikman}},
  \bibinfo{journal}{Science} \textbf{\bibinfo{volume}{27}},
  \bibinfo{pages}{779} (\bibinfo{year}{2000}).

\bibitem[{\citenamefont{Silvestrov and Imry}(2003)}]{silvestrov03}
\bibinfo{author}{\bibfnamefont{P.~G.} \bibnamefont{Silvestrov}}
  \bibnamefont{and} \bibinfo{author}{\bibfnamefont{Y.}~\bibnamefont{Imry}},
  \bibinfo{journal}{Phys.\ Rev.\ Lett.} \textbf{\bibinfo{volume}{90}},
  \bibinfo{pages}{106602} (\bibinfo{year}{2003}).

\bibitem[{\citenamefont{Oreg and Gefen}(1997)}]{oreg97}
\bibinfo{author}{\bibfnamefont{Y.}~\bibnamefont{Oreg}} \bibnamefont{and}
  \bibinfo{author}{\bibfnamefont{Y.}~\bibnamefont{Gefen}},
  \bibinfo{journal}{Phys.\ Rev.\ B} \textbf{\bibinfo{volume}{55}},
  \bibinfo{pages}{13726} (\bibinfo{year}{1997}).

\bibitem[{\citenamefont{Yeyati and Buttiker}(2000)}]{levyyeyati00}
\bibinfo{author}{\bibfnamefont{A.~L.} \bibnamefont{Yeyati}} \bibnamefont{and}
  \bibinfo{author}{\bibfnamefont{M.}~\bibnamefont{Buttiker}},
  \bibinfo{journal}{Phys.\ Rev.\ B} \textbf{\bibinfo{volume}{62}},
  \bibinfo{pages}{7307} (\bibinfo{year}{2000}).

\bibitem[{\citenamefont{Xu and Gu}(2001)}]{xu01}
\bibinfo{author}{\bibfnamefont{H.~Q.} \bibnamefont{Xu}} \bibnamefont{and}
  \bibinfo{author}{\bibfnamefont{B.-Y.} \bibnamefont{Gu}},
  \bibinfo{journal}{J.\ Phys.: Condens.\ Matter} \textbf{\bibinfo{volume}{13}},
  \bibinfo{pages}{3599} (\bibinfo{year}{2001}).

\bibitem[{\citenamefont{Silvestrov and Imry}(2000)}]{silvestrov00}
\bibinfo{author}{\bibfnamefont{P.~G.} \bibnamefont{Silvestrov}}
  \bibnamefont{and} \bibinfo{author}{\bibfnamefont{Y.}~\bibnamefont{Imry}},
  \bibinfo{journal}{Phys.\ Rev.\ Lett.} \textbf{\bibinfo{volume}{85}},
  \bibinfo{pages}{2565} (\bibinfo{year}{2000}).

\bibitem[{hac()}]{hackenbroich01}
\bibinfo{note}{For a rewiev see G. Hackenbroich, Phys. Rep. {\bf 343}, 463
  (2001).}

\bibitem[{\citenamefont{Nockel and Stone}(1994)}]{nockel94}
\bibinfo{author}{\bibfnamefont{J.~U.} \bibnamefont{Nockel}} \bibnamefont{and}
  \bibinfo{author}{\bibfnamefont{A.~D.} \bibnamefont{Stone}},
  \bibinfo{journal}{Phys.\ Rev. B} \textbf{\bibinfo{volume}{50}},
  \bibinfo{pages}{17415} (\bibinfo{year}{1994}).

\bibitem[{\citenamefont{Lent and Kirkner}(1990)}]{kirknerqtbm}
\bibinfo{author}{\bibfnamefont{C.~S.} \bibnamefont{Lent}} \bibnamefont{and}
  \bibinfo{author}{\bibfnamefont{D.~J.} \bibnamefont{Kirkner}},
  \bibinfo{journal}{J.\ Appl.\ Phys.} \textbf{\bibinfo{volume}{67}},
  \bibinfo{pages}{6353} (\bibinfo{year}{1990}).

\bibitem[{not({\natexlab{a}})}]{nota1}
\bibinfo{note}{Although the width and energies of the resonances vary with $d$
  and $\lambda_D$, we found that the qualitative behavior of the transmission
  amplitude is not affected by the choiche of those parameters.}

\bibitem[{not({\natexlab{b}})}]{nota4}
\bibinfo{note}{In addition, due to the very small energy of the incoming
  electron which is comparable to the numerical error in the determination of
  the energy of the QBLs, the energy where an additional electron gets trapped
  into the QD is fixed in the middle of the first two QBLs rather than from
  comparison to the Fermi energy.}

\bibitem[{not({\natexlab{c}})}]{notaxu}
\bibinfo{note}{While in our case the phase drops happen as the occupancy of the
  dot changes by one unit and do not correspond necessarily to a zero of the
  transmission, in Ref.~\onlinecite{xu01}, the drops always coincide with a
  transmission zero. This is due to the constraint of an integer occupancy in
  our electrostatic model.}

\bibitem[{\citenamefont{Fogler}(2005)}]{fogler05}
\bibinfo{author}{\bibfnamefont{M.~M.} \bibnamefont{Fogler}},
  \bibinfo{journal}{Phys.\ Rev.\ Lett.} \textbf{\bibinfo{volume}{94}},
  \bibinfo{pages}{56405} (\bibinfo{year}{2005}).

\bibitem[{\citenamefont{Bertoni and Goldoni}(2006)}]{bertoni_jce}
\bibinfo{author}{\bibfnamefont{A.}~\bibnamefont{Bertoni}} \bibnamefont{and}
  \bibinfo{author}{\bibfnamefont{G.}~\bibnamefont{Goldoni}},
  \bibinfo{journal}{J.\ Comp.\ Electron.} \textbf{\bibinfo{volume}{5}},
  \bibinfo{pages}{177} (\bibinfo{year}{2006}).

\bibitem[{\citenamefont{Bertoni and Goldoni}(2007)}]{bertoni_unpub}
\bibinfo{author}{\bibfnamefont{A.}~\bibnamefont{Bertoni}} \bibnamefont{and}
  \bibinfo{author}{\bibfnamefont{G.}~\bibnamefont{Goldoni}}
  (\bibinfo{year}{2007}), \bibinfo{note}{(unpublished)}.

\bibitem[{not({\natexlab{d}})}]{nota5}
\bibinfo{note}{The approach described replicate the ``quantum transmitting
  boundary method'' of Ref.~\onlinecite{kirknerqtbm} and thereafter
  applied to the modeling of many micro- and nano-electronic
  systems. In fact, in the original formulation, the problem was the
  solution of a single-particle 2D Schr{\"o}dinger equation with open
  boundaries, and its extension to three (and conceptually even more)
  dimensions was straightforward. By considering that the equation for
  a single particle in N dimensions and that for N particles in
  one dimension have the same form, it is possible to take profit of the
  technique developed for the former in order to solve the latter.}

\bibitem[{\citenamefont{Fano}(1961)}]{fano61}
\bibinfo{author}{\bibfnamefont{U.}~\bibnamefont{Fano}},
  \bibinfo{journal}{Phys.\ Rev.} \textbf{\bibinfo{volume}{124}},
  \bibinfo{pages}{1866} (\bibinfo{year}{1961}).

\bibitem[{\citenamefont{Taniguchi and B{\"u}ttiker}(1999)}]{taniguchi99}
\bibinfo{author}{\bibfnamefont{T.}~\bibnamefont{Taniguchi}} \bibnamefont{and}
  \bibinfo{author}{\bibfnamefont{M.}~\bibnamefont{B{\"u}ttiker}},
  \bibinfo{journal}{J.\ Phys.: Condens.\ Matter} \textbf{\bibinfo{volume}{60}},
  \bibinfo{pages}{13814} (\bibinfo{year}{1999}).

\bibitem[{\citenamefont{Rontani}(2006)}]{rontani06}
\bibinfo{author}{\bibfnamefont{M.}~\bibnamefont{Rontani}},
  \bibinfo{journal}{Phys.\ Rev.\ Lett.} \textbf{\bibinfo{volume}{97}},
  \bibinfo{pages}{76801} (\bibinfo{year}{2006}).

\bibitem[{\citenamefont{Tekman and Bagwell}(1993)}]{tekman93}
\bibinfo{author}{\bibfnamefont{E.}~\bibnamefont{Tekman}} \bibnamefont{and}
  \bibinfo{author}{\bibfnamefont{P.~F.} \bibnamefont{Bagwell}},
  \bibinfo{journal}{Phys.\ Rev. B} \textbf{\bibinfo{volume}{48}},
  \bibinfo{pages}{2553} (\bibinfo{year}{1993}).

\bibitem[{\citenamefont{Racec and Wulf}(2001)}]{racec01}
\bibinfo{author}{\bibfnamefont{E.~R.} \bibnamefont{Racec}} \bibnamefont{and}
  \bibinfo{author}{\bibfnamefont{U.}~\bibnamefont{Wulf}},
  \bibinfo{journal}{Phys.\ Rev. B} \textbf{\bibinfo{volume}{64}},
  \bibinfo{pages}{115318} (\bibinfo{year}{2001}).

\bibitem[{\citenamefont{Entin-Wohlman et~al.}(2002)\citenamefont{Entin-Wohlman,
  Aharony, Imry, and Levinson}}]{wohlman02}
\bibinfo{author}{\bibfnamefont{O.}~\bibnamefont{Entin-Wohlman}},
  \bibinfo{author}{\bibfnamefont{A.}~\bibnamefont{Aharony}},
  \bibinfo{author}{\bibfnamefont{Y.}~\bibnamefont{Imry}}, \bibnamefont{and}
  \bibinfo{author}{\bibfnamefont{Y.}~\bibnamefont{Levinson}},
  \bibinfo{journal}{J.\ Low Temp.\ Phys.} \textbf{\bibinfo{volume}{126}},
  \bibinfo{pages}{1251} (\bibinfo{year}{2002}).

\bibitem[{\citenamefont{Fuhrer et~al.}(2006)\citenamefont{Fuhrer, Brusheim,
  Ihn, Sigrist, Ensslin, Wegscheider, and Bichler}}]{fuhrer06}
\bibinfo{author}{\bibfnamefont{A.}~\bibnamefont{Fuhrer}},
  \bibinfo{author}{\bibfnamefont{P.}~\bibnamefont{Brusheim}},
  \bibinfo{author}{\bibfnamefont{T.}~\bibnamefont{Ihn}},
  \bibinfo{author}{\bibfnamefont{M.}~\bibnamefont{Sigrist}},
  \bibinfo{author}{\bibfnamefont{K.}~\bibnamefont{Ensslin}},
  \bibinfo{author}{\bibfnamefont{W.}~\bibnamefont{Wegscheider}},
  \bibnamefont{and} \bibinfo{author}{\bibfnamefont{M.}~\bibnamefont{Bichler}},
  \bibinfo{journal}{Phys.\ Rev.\ B} \textbf{\bibinfo{volume}{73}},
  \bibinfo{pages}{205326} (\bibinfo{year}{2006}).

\bibitem[{\citenamefont{Karrasch et~al.}(2007)\citenamefont{Karrasch, Hecht,
  Weichselbaum, Oreg, von Delft, , and Meden}}]{karrasch07}
\bibinfo{author}{\bibfnamefont{C.}~\bibnamefont{Karrasch}},
  \bibinfo{author}{\bibfnamefont{T.}~\bibnamefont{Hecht}},
  \bibinfo{author}{\bibfnamefont{A.}~\bibnamefont{Weichselbaum}},
  \bibinfo{author}{\bibfnamefont{Y.}~\bibnamefont{Oreg}},
  \bibinfo{author}{\bibfnamefont{J.}~\bibnamefont{von Delft}}, ,
  \bibnamefont{and} \bibinfo{author}{\bibfnamefont{V.}~\bibnamefont{Meden}},
  \bibinfo{journal}{Phys.\ Rev.\ Lett.} \textbf{\bibinfo{volume}{98}},
  \bibinfo{pages}{186802} (\bibinfo{year}{2007}).

\end{thebibliography}
\end{document}